\def\logoname{logo}

\documentclass[journal,twoside,web]{ieeecolor}

\usepackage{soul}
\usepackage{generic}
\usepackage{cite}
\usepackage{amsmath,amssymb,amsfonts}
\usepackage{algorithmic}
\usepackage{graphicx}
\usepackage{algorithm,algorithmic}
\usepackage{booktabs}
\usepackage{textcomp}
\usepackage{caption}
\usepackage{multirow}
\usepackage{orcidlink}

\def\BibTeX{{\rm B\kern-.05em{\sc i\kern-.025em b}\kern-.08em
    T\kern-.1667em\lower.7ex\hbox{E}\kern-.125emX}}
\usepackage{xcolor}
\usepackage{colortbl}
\hypersetup{
    colorlinks=true,
    linkcolor=blue,
    citecolor=blue,
    urlcolor=blue
}

\usepackage{tgpagella}

\begin{document}
\title{Expert-Guided Explainable Few-Shot Learning with Active Sample Selection for Medical Image Analysis}
\author{Longwei Wang\orcidlink{0009-0002-0638-5637}, \IEEEmembership{Member, IEEE}, Ifrat Ikhtear Uddin\orcidlink{0009-0004-4378-1051}, and KC Santosh\orcidlink{0000-0003-4176-0236}, \IEEEmembership{Senior Member, IEEE}
\thanks{Longwei Wang, Ifrat Ikhtear Uddin, and KC Santosh are with the AI Research, Department of Computer Science at University of South Dakota, Vermillion, SD 57069, USA (e-mail: longwei.wang@usd.edu; ifratikhtear.uddin@coyotes.usd.edu; kc.santosh@usd.edu)}%
}
\maketitle

\begin{abstract}
Medical image analysis faces two critical challenges: scarcity of labeled data and lack of model interpretability, both hindering clinical AI deployment. Few-shot learning (FSL) addresses data limitations but lacks transparency in predictions. Active learning (AL) methods optimize data acquisition but overlook interpretability of acquired samples. We propose a dual-framework solution: Expert-Guided Explainable Few-Shot Learning (EGxFSL) and Explainability-Guided AL (xGAL). EGxFSL integrates radiologist-defined regions-of-interest as spatial supervision via Grad-CAM-based Dice loss, jointly optimized with prototypical classification for interpretable few-shot learning. xGAL introduces iterative sample acquisition prioritizing both predictive uncertainty and attention misalignment, creating a closed-loop framework where explainability guides training and sample selection synergistically. 
On the BraTS (MRI), VinDr-CXR (chest X-ray), and SIIM-COVID-19 (chest X-ray) datasets, we achieve accuracies of 92\%, 76\%, and 62\%, respectively, consistently outperforming non-guided baselines across all datasets. Under severe data constraints, xGAL achieves 76\% accuracy with only 680 samples versus 57\% for random sampling. Grad-CAM visualizations demonstrate guided models focus on diagnostically relevant regions, with generalization validated on breast ultrasound confirming cross-modality applicability.

\end{abstract}


\section{Introduction}
\label{sec:introduction}

\IEEEPARstart{D}{eep} learning has achieved remarkable progress in medical image analysis, powering applications such as disease detection, anatomical segmentation, and abnormality classification across a variety of imaging modalities including MRI, CT, and chest X-ray ~\cite{krizhevsky2012imagenet,chen2025survey}.  However, these advances rely heavily on the availability of large-scale, high-quality annotated datasets. Such resources are particularly scarce in clinical domains due to the time-consuming and expensive nature of expert labeling. In many practical scenarios, especially involving rare diseases or resource-constrained healthcare settings, only a few labeled examples per diagnostic category are available. This limitation significantly impairs the generalization ability of standard supervised deep learning models and presents a major bottleneck for clinical deployment.

Few-shot learning (FSL) offers a promising solution to this data scarcity by enabling models to generalize from only a handful of labeled instances per class~\cite{snell2017prototypical}. FSL methods, such as prototypical networks, leverage metric-based learning to compute class prototypes in an embedding space~\cite{laenen2021episodes,song2025interactive}. Despite their data efficiency, most FSL models lack interpretability, rendering their decision-making process opaque to clinicians~\cite{mumuni2024improving,liu2025envisioning}. In high-stakes medical environments, where trust in AI predictions is critical, this lack of transparency can be a dealbreaker.

Active learning (AL) complements FSL by minimizing annotation costs through intelligent sample selection and model fine-tuning~\cite{wu2023improved,hubotter2024active}. Deep active learning (DeepAL) combines AL strategies with deep networks to iteratively identify the most informative unlabeled samples for annotation~\cite{ren2021surveydeepactivelearning,aghdam2019active}. Nevertheless, traditional AL methods frequently rely on softmax-based uncertainty metrics, which are known to be unreliable in high-dimensional and overconfident neural models~\cite{wang2017cost}. Furthermore, conventional AL strategies focus solely on improving predictive performance.  They do not consider whether selected samples contribute to model interpretability -- a critical requirement in clinical contexts~\cite{zhang2025uncertainty}.

Meanwhile, explainable artificial intelligence (xAI) has emerged as a vital research direction to address the transparency limitations of deep models. Post-hoc attribution techniques such as Grad-CAM~\cite{selvaraju2017grad} and SHAP~\cite{lundberg2017unified} visualize salient input regions that influence a model's prediction, offering insights into its decision-making. However, these explanations are usually generated after model training and do not influence the learning process itself. Consequently, even visually plausible explanations may not correspond to clinically meaningful features, as models can still rely on spurious correlations.

To address these challenges, we propose a unified dual-framework approach that bridges the gap between interpretability and data efficiency in few-shot medical image classification. Our contribution consists of two complementary components: Expert-Guided Explainable Few-Shot Learning (EGxFSL) and Explainability-Guided Active Learning (xGAL). The EGxFSL framework integrates expert-provided annotations during training to ensure clinically-aligned model attention.  xGAL guides the sample acquisition process through attribution-based reasoning to maximize data efficiency. Critically, xGAL leverages the explainability capabilities developed in EGxFSL during its iterative refinement process, creating a unified system where interpretability guides data acquisition and newly acquired expert knowledge continuously enhances model explainability. This symbiotic relationship ensures that learned representations are both diagnostically relevant and data-efficient, satisfying the twin imperatives of performance and transparency.

We conduct comprehensive experiments on three expert-annotated medical imaging datasets: BraTS (multi-modal MRI scans of brain tumors~\cite{menze2014multimodal}), VinDr-CXR (large-scale chest X-ray dataset with thoracic disease findings~\cite{nguyen2022vindr}), and SIIM-FISABIO-RSNA-COVID-19~\cite{lakhani2021siim}. Empirical results demonstrate that both EGxFSL and xGAL outperform baseline FSL models and traditional AL approaches in classification accuracy and interpretability metrics, while significantly reducing the amount of labeled data required. Our contributions are:

\begin{enumerate}
    \item We propose Expert-Guided Explainable Few-Shot Learning (EGxFSL), which integrates radiologist-provided ROIs into training via a Dice-based explanation loss that aligns Grad-CAM heatmaps with expert annotations alongside the classification objective.
    
    \item We introduce Explainability-Guided Active Learning (xGAL), a sample acquisition strategy that jointly considers predictive uncertainty and explanation misalignment, using EGxFSL as the core learner to improve both accuracy and interpretability.

    \item We validate the unified framework on three datasets across two imaging modalities, demonstrating consistent improvements in performance and attribution quality under extreme data scarcity.
\end{enumerate}

\begin{figure*}[htb]
    \centering
    \includegraphics[width=0.90\textwidth]{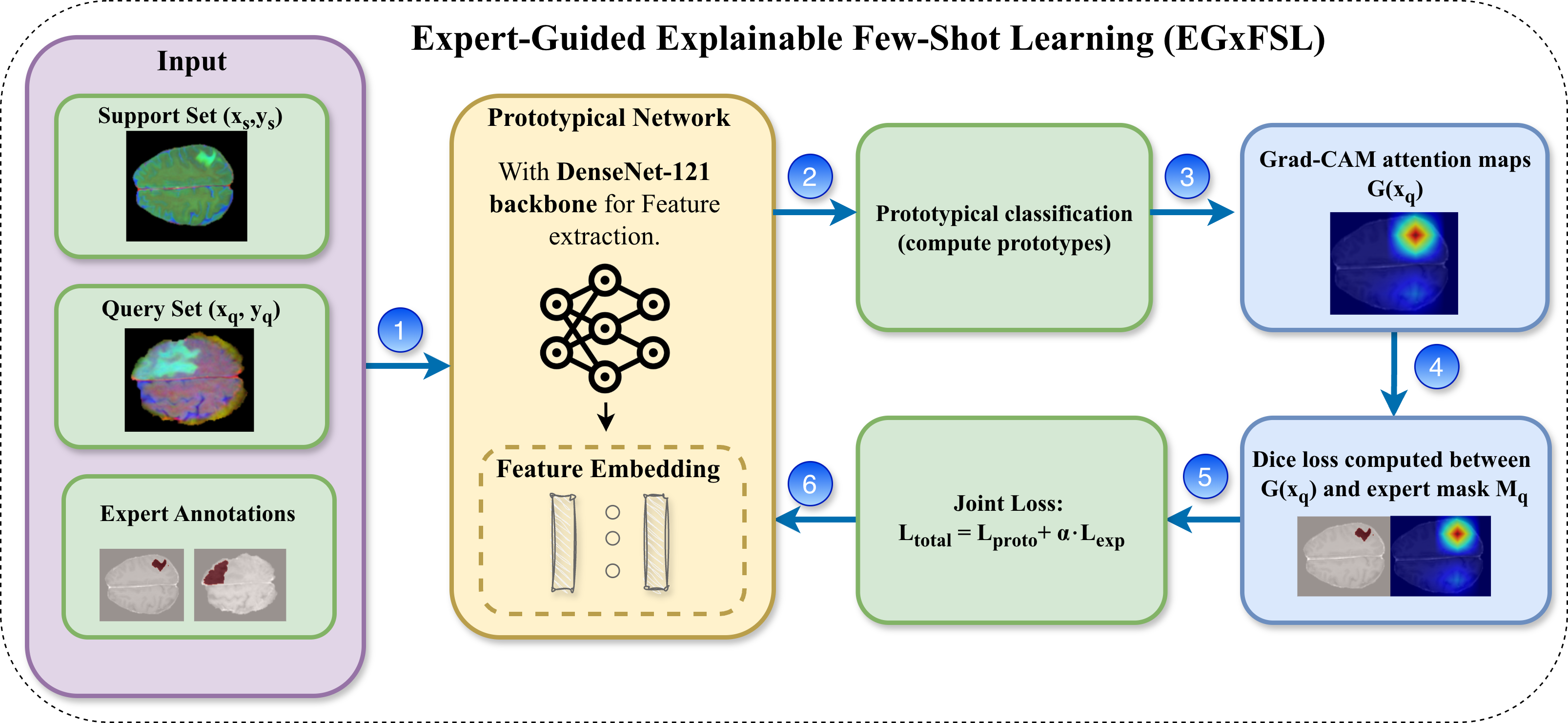}
    \caption{Expert-Guided Explainable Few-Shot Learning (EGxFSL) Framework. The pipeline consists of six key steps: (1) Input processing where support sets ($x_s, y_s$) and query samples ($x_q, y_q$) are fed into the system, with expert annotations available for both support and query sets. (2) Feature extraction using a DenseNet-121 backbone within the prototypical network to generate embeddings in a learned embedding space. (3) Prototypical classification where class prototypes are computed from support embeddings and used to classify query samples; (4) GradCAM attention map generation ($G(x_q)$) for query samples to visualize model focus regions; (5) Explainability alignment where Dice loss is computed between the generated GradCAM attention maps and expert annotations from query samples ($M_q$) to measure alignment; (6) Joint optimization using the combined loss function $\mathcal{L}_{total}$ (Equation~\ref{eq:total_loss}) to simultaneously train for accurate classification and clinically-aligned attention for trustworthy, explainable diagnosis.}
    \label{fig:exp-guided}
\end{figure*}

\section{Related Work}
\label{sec:related_work}

\subsection{FSL in Medical Imaging}

FSL has emerged as a compelling solution to the widespread issue of limited labeled data in medical imaging. Obtaining expert annotations is both labor-intensive and cost-prohibitive. The core idea behind FSL is to enable models to generalize from a small number of labeled instances per class, often using meta-learning or metric-based strategies. One of the most widely adopted frameworks in this space is the prototypical network~\cite{snell2017prototypical}, which learns a metric space where classification is performed by computing distances to class prototypes formed from support examples.

In the medical domain, FSL has shown promise in tasks such as histopathology image classification~\cite{quan2024dual}, skin lesion diagnosis, and diabetic retinopathy detection. Data scarcity is especially pronounced in these areas due to disease rarity or privacy constraints~\cite{hussain2025few}. Ouahab and Ahmed~\cite{ouahab2025protomed} proposed ProtoMed, a regularized version of prototypical networks tailored for medical imaging, which incorporates auxiliary information to stabilize learning in low-data regimes. Despite these advancements, the majority of FSL models prioritize classification accuracy while treating model interpretability as secondary. This shortcoming is particularly problematic in healthcare, where clinical decisions must be explainable and justifiable.

\subsection{AL for Annotation Efficiency}

AL has long been recognized as an effective strategy for reducing annotation burdens by selectively querying the most informative unlabeled samples for labeling~\cite{hsu2015active,ho2024learning,nayyem2024bridging,wang2024enhanced,sung2018learning,ravi2017optimization}. In medical imaging, AL is especially useful due to the high cost and limited availability of expert annotators~\cite{biswas2023active}. Recent methods such as ACFT (Active, Continual Fine-Tuning)~\cite{zhou2021active} have successfully integrated active sampling into deep learning workflows. ACFT employs entropy and diversity-based acquisition functions to select diverse and uncertain samples across multiple clinical tasks, including colonoscopy frame classification and pulmonary embolism detection. Similarly, Hao et al.~\cite{hao2021transfer} combined uncertainty sampling with a query-by-committee strategy to construct a transfer learning-based AL pipeline for brain tumor classification.

 While these methods significantly reduce labeling costs and improve generalization, they primarily rely on uncertainty estimation through softmax probabilities or ensemble disagreement. These uncertainty metrics are known to be overconfident and unreliable in deep models~\cite{wang2017cost}. Moreover, these acquisition strategies do not incorporate any measure of model interpretability or alignment with domain knowledge. This limitation reduces their usefulness in high-stakes clinical settings where human oversight and trust are essential.

\subsection{xAI in Medical Imaging}

 xAI has become a critical focus in medical AI to ensure that model decisions are transparent, justifiable, and aligned with human expert reasoning. Among the most popular post-hoc explanation methods are Grad-CAM~\cite{selvaraju2017grad} and SHAP~\cite{lundberg2017unified}, which visualize salient regions of input images that contribute to model predictions. These methods have been extensively applied in medical tasks such as diabetic retinopathy detection, chest X-ray diagnosis, and brain tumor segmentation~\cite{loh2022application,wang2019representation,wang2025bridging,chataut2024shape,wang2025explainability,wang2021explaining,sadeghi2024review}.

However, a key limitation of most xAI techniques is that they are applied after model training. This means they do not influence the model's learning trajectory. As a result, models may still rely on spurious or non-causal features during inference, even if their explanations appear visually plausible. This disconnect between training and interpretation reduces the clinical reliability of post-hoc explanations and undermines efforts to build transparent AI systems.

\subsection{Explanation-Guided Learning and Supervised Attention Alignment}

To address the limitations of post-hoc interpretability, recent works have proposed integrating explanation guidance directly into the model training process. For example, Šefčík et al.~\cite{vsefvcik2023improving} used Layer-wise Relevance Propagation (LRP) to guide model attention towards glioma tumor regions in brain MRI. Caragliano et al.~\cite{caragliano2025doctorintheloopexplainablemultiviewdeep} introduced the Doctor-in-the-Loop framework, which incorporates radiologist-annotated regions-of-interest (ROIs) to guide model learning in non-small cell lung cancer CT imaging. These approaches demonstrate the potential of explanation-supervised learning in improving spatial fidelity of model attention and enhancing clinical trust. Sun et al.~\cite{sun2021explanation} applied explanation-guided training to cross-domain few-shot image classification in non-medical domains.  However, these efforts are generally confined to fully supervised settings or natural image domains. They do not address the combined challenges of FSL, sample efficiency, and clinical interpretability. Crucially, there remains a gap in unifying explanation-guided training with AL for interpretable model refinement under data-scarce conditions.

Despite progress in FSL, AL, and xAI, their integration remains unexplored in medical imaging. Existing FSL models lack spatial interpretability, AL ignores clinical alignment, and explanation supervision is rarely used for data acquisition. We address these limitations through a dual-framework: expert-guided FSL that integrates radiologist ROIs via Grad-CAM-based Dice loss, and explainability-aware AL that selects samples using predictive uncertainty and attention misalignment, enabling data-efficient learning with high interpretability.

\section{Methodology}
\label{sec:methodology}
We propose a dual-framework integrating radiologist-provided spatial annotations into both model training and sample selection to ensure learned representations are accurate and clinically meaningful.


\subsection{EGxFSL: Expert-Guided Explainable Few-Shot Learning}
\label{sec:EGxFSL}

We address few-shot classification using prototypical networks, which compare query images to class prototypes in a learned metric space. Our EGxFSL framework ensures models focus on diagnostically relevant features through Grad-CAM-based attention supervision, guiding learning toward clinically meaningful regions rather than spurious correlations. This makes model decisions interpretable—when Grad-CAM shows the model focuses on the same anatomical structures experts use for diagnosis, clinicians can trust its predictions.

\subsubsection{Prototypical Network Foundation}

We adopt the prototypical network~\cite{snell2017prototypical,pahde2021multimodal} as our base FSL architecture. In an $N$-way $K$-shot classification setup, the support set \( S = \{(x_s, y_s)\}_{s=1}^{N \cdot K} \) contains \( K \) labeled examples from each of the \( N \) classes, while the query set \( Q = \{(x_q, y_q)\}_{q=1}^{N \cdot Q} \) is used for evaluation.

A feature extractor \( f_{\theta}(x) \), implemented as a DenseNet-121 backbone with the classification head removed, maps input images to a $d$-dimensional embedding space. For each class \( k \), we compute a prototype vector \( c_k \) as the mean embedding of all support samples belonging to that class:
\begin{equation}
c_k = \frac{1}{|S_k|} \sum_{(x_i, y_i) \in S_k} f_{\theta}(x_i).
\end{equation}
Classification of a query image \( x_q \) is performed by computing distances to all class prototypes and applying softmax:
\begin{equation}
p(y=k \mid x_q) = \frac{\exp(-\|f_{\theta}(x_q) - c_k\|^2)}{\sum_{k'} \exp(-\|f_{\theta}(x_q) - c_{k'}\|^2)}.
\end{equation}
The standard few-shot classification loss is the negative log-likelihood:
\begin{equation}
\mathcal{L}_{\text{proto}} = -\log p(y = y_q \mid x_q).
\label{eq:proto_loss}
\end{equation}

\subsubsection{Expert-Guided Attention Alignment via Grad-CAM}

To ensure the model focuses on diagnostically meaningful regions rather than spurious correlations, we incorporate radiologist-provided ROIs into the training objective. The key idea is to penalize the model when its internal attention diverges from expert annotations, steering it toward clinically relevant features.

We employ Grad-CAM~\cite{selvaraju2017grad} to generate class-discriminative localization maps $G$ that visualize which image regions the model uses for predictions. For a given query image $x_q$, we compare the model's attention map $G$ against the corresponding expert-provided binary mask $M$ indicating the ground truth abnormality location. We define an explanation alignment loss using the Dice similarity coefficient:
\begin{equation}
\mathcal{L}_{\text{exp}} = 1 - \frac{2 \cdot |G \cap M|}{|G| + |M|}.
\label{eq:exp_loss}
\end{equation}
This loss promotes spatial alignment between model attention and clinically relevant regions. When the model correctly focuses on tumor regions annotated by radiologists, $\mathcal{L}_{\text{exp}}$ is low; when attention is misaligned, the loss increases, providing gradient signals to correct the model's focus during backpropagation.

\subsubsection{Joint Optimization Objective}

We integrate the classification and explanation objectives into a unified training loss:
\begin{equation}
\mathcal{L}_{\text{total}} = \mathcal{L}_{\text{proto}} + \alpha \cdot \mathcal{L}_{\text{exp}},
\label{eq:total_loss}
\end{equation}
where $\alpha$ is a hyperparameter controlling the strength of explanation-based supervision.  Minimizing $\mathcal{L}_{\text{total}}$ encourages the model to achieve both high classification accuracy and clinically-aligned attention patterns. As illustrated in Figure~\ref{fig:exp-guided}, our EGxFSL framework produces models that are both accurate and interpretable, particularly important in low-data environments where overfitting to irrelevant features is a common risk.

\subsection{ xGAL: Explainability-Guided Active Learning}
\label{sec:active_learning}

xGAL addresses scenarios where acquiring even limited labeled data is prohibitively expensive. The framework identifies which unlabeled samples would most improve both model performance and interpretability if added to the training set, combining predictive uncertainty with attention alignment unlike traditional AL methods that use uncertainty alone.

Regarding scope and applicability, our xGAL framework applies to scenarios where spatial annotations exist in public datasets (BraTS, VinDr-CXR, SIIM-COVID) but strategic sample selection optimizes which images to include under limited training budgets. This addresses efficient utilization of existing annotated repositories particularly relevant for adapting models to new clinical sites or specialized protocols using minimal data from large pre-annotated collections.

\begin{figure*}[tpb]
    \centering
    \includegraphics[width=0.90\textwidth]{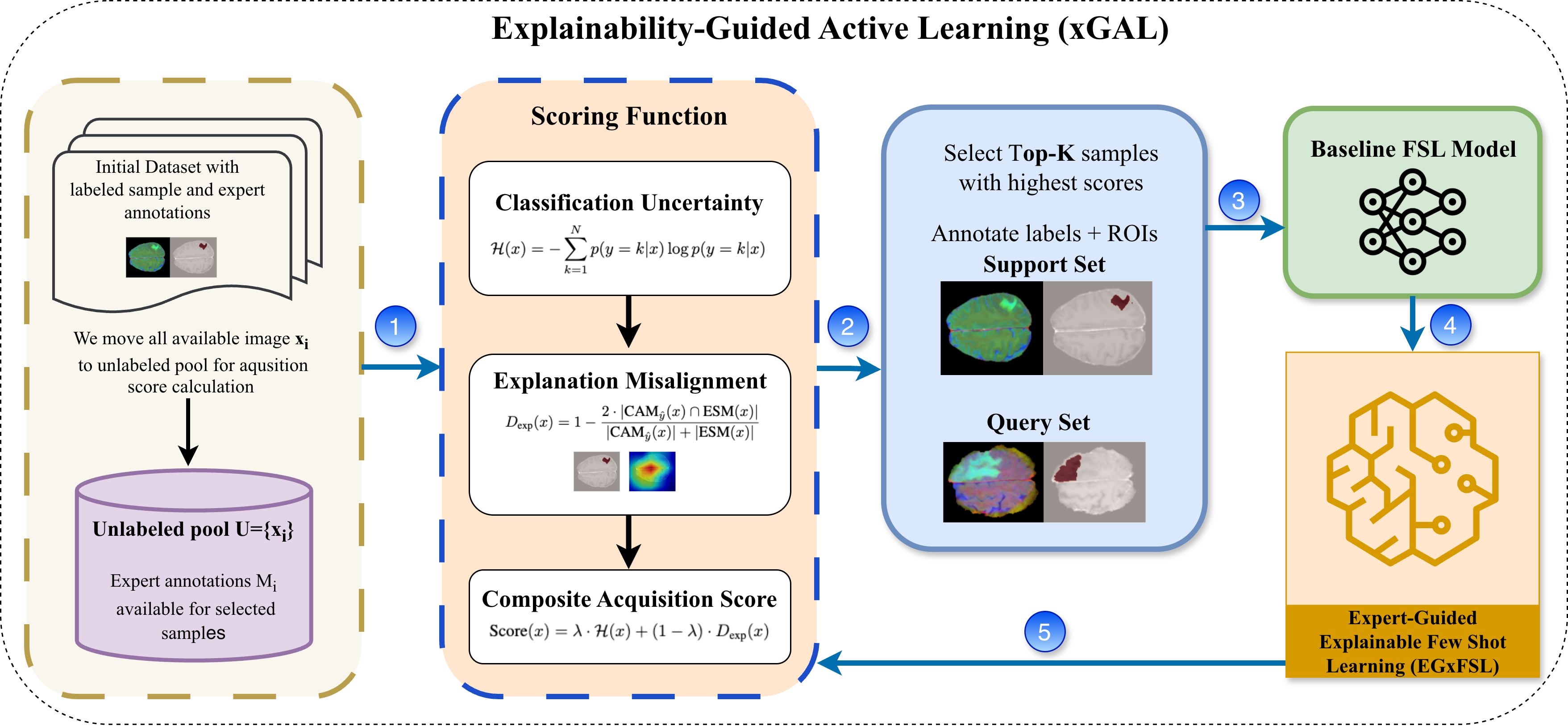}
    \caption{Explainability-Guided Active Learning (xGAL) Framework. The framework operates in an iterative cycle with five key components: (1) Starting with an unlabeled pool $U=\{x_i\}$ where expert annotations are available for selected samples; (2) A composite scoring function that combines classification uncertainty $\mathcal{H}(x)$ and explanation misalignment $D_{\text{exp}}(x)$; (3) Composite acquisition score $\text{Score}(x)$ where $\lambda$ balances uncertainty and misalignment; (4) Selection of top-K samples with highest composite scores, followed by expert annotation to obtain both class labels and diagnostic ROI masks for support and query sets; (5) Model retraining using the EGxFSL framework (Figure~\ref{fig:exp-guided}). This iterative system progressively improves both classification accuracy and explanation quality with minimal labeled data.}
    \label{fig:active-learning-frm}
\end{figure*}

\subsubsection{Scoring Mechanism for Sample Acquisition}
Let \( \mathcal{U} \) denote the pool of unlabeled images. For each sample \( x \in \mathcal{U} \), we compute an acquisition score combining two criteria:

\smallskip
\noindent {\em a) Classification Uncertainty.} We quantify predictive uncertainty using Shannon entropy over the model's softmax output:
\begin{equation}
\mathcal{H}(x) = - \sum_{k=1}^{N} p(y=k|x) \log p(y=k|x),
\end{equation}
where \( p(y=k|x) \) denotes the probability of class \( k \) under the current FSL model.  High entropy indicates the model is uncertain about the correct class, suggesting this sample could provide valuable learning signal if labeled.

\smallskip
\noindent {\em b) Explanation Misalignment.} Using Grad-CAM, we generate a class activation map \( \text{CAM}_{\hat{y}}(x) \) for the predicted class \( \hat{y} = \arg\max_k p(y=k|x) \). We compare this attention map to the expert-supplied spatial annotation \( \text{ESM}(x) \), which captures the expected diagnostic region. The discrepancy is measured using Dice-based loss:
\begin{equation}
D_{\text{exp}}(x) = 1 - \frac{2 \cdot |\text{CAM}_{\hat{y}}(x) \cap \text{ESM}(x)|}{|\text{CAM}_{\hat{y}}(x)| + |\text{ESM}(x)|}.
\end{equation}
 High misalignment indicates the model's attention is not focused on clinically relevant regions, suggesting this sample could help correct the model's spatial reasoning if added to the training set.

\smallskip

\noindent {\em c) Composite Acquisition Score.} The overall acquisition score balances both criteria:
\begin{equation}
\text{Score}(x) = \lambda \cdot \mathcal{H}(x) + (1 - \lambda) \cdot D_{\text{exp}}(x),
\label{eq:acc_score}
\end{equation}
where \( \lambda \in [0,1] \) controls the trade-off between uncertainty and interpretability. Through empirical validation, we find that $\lambda = 0.5$ for BraTS and SIIM-COVID, and $\lambda = 0.6$ for VinDr-CXR provide robust performance, with optimal values depending on dataset characteristics such as class balance and annotation quality.

\subsubsection{Active Learning Procedure}

At each acquisition round, we select a batch \( \mathcal{B} \subset \mathcal{U} \) of the top-\( K \) unlabeled samples with the highest scores according to Equation~\eqref{eq:acc_score}. For each selected sample, we retrieve: (1) the ground truth class label \( y \); and (2) the corresponding expert-provided spatial mask \( \text{ESM}(x) \), either as a segmentation map (BraTS) or a binary mask derived from bounding boxes (VinDr-CXR, SIIM-COVID).

The selected samples are added to the labeled support set \( \mathcal{S} \), and the model is fine-tuned using our EGxFSL framework (Section~\ref{sec:EGxFSL}). This cycle repeats iteratively, with each round retraining the model on the progressively expanded labeled set.  Our experiments demonstrate that this strategic refinement yields substantial improvements in both accuracy and attention alignment, even when the total number of labeled samples is limited to 680 across all classes.

\subsubsection{Algorithm Summary}

Algorithm~\ref{alg:active_learning} summarizes the complete xGAL procedure, which iteratively refines the model using strategically selected samples that improve both classification performance and interpretability.

\begin{algorithm}[tbp]
\caption{Explainability-Guided Active Learning (xGAL) Framework}
\label{alg:active_learning}
\begin{algorithmic}[1]
\REQUIRE Unlabeled pool \( \mathcal{U} \), current model \( f_{\theta} \), batch size \( K \), dataset annotations \( \{\text{ESM}(x) : x \in \mathcal{U}\} \)
\FOR{each \( x \in \mathcal{U} \)}
    \STATE Compute classification entropy \( \mathcal{H}(x) \)
    \STATE Generate Grad-CAM heatmap \( \text{CAM}_{\hat{y}}(x) \)
    \STATE Retrieve expert annotation \( \text{ESM}(x) \)
    \STATE Compute explanation misalignment \( D_{\text{exp}}(x) \) using \( \text{CAM}_{\hat{y}}(x) \) and \( \text{ESM}(x) \)
    \STATE Compute score: \( \text{Score}(x) = \lambda \mathcal{H}(x) + (1-\lambda) D_{\text{exp}}(x) \)
\ENDFOR
\STATE Select top-\( K \) samples: \( \mathcal{B} \leftarrow \text{TopK}_{x \in \mathcal{U}} (\text{Score}(x)) \)
\STATE Add selected samples with their existing labels and annotations to \( \mathcal{S} \)
\STATE Fine-tune model on \( \mathcal{S} \) using \( \mathcal{L}_{\text{total}} \)
\STATE Remove \( \mathcal{B} \) from \( \mathcal{U} \)
\end{algorithmic}
\end{algorithm}

\subsection{Hyperparameter Selection Guidelines}
\label{sec:hyperparameter_selection}

Our framework introduces two key hyperparameters: $\alpha$ (explanation supervision strength) and $\lambda$ (uncertainty-interpretability balance). Based on comprehensive sensitivity analysis across datasets, we provide systematic selection guidelines:

\smallskip

\noindent {\em a) Explanation Weight ($\alpha$).} Optimal performance is achieved at $\alpha = 0.10$ across all datasets (Figure~\ref{fig:alpvavsacc}). We recommend $\alpha = 0.10$ as default, increasing to 0.20-0.30 if attention alignment is poor, or decreasing to 0.05 if classification accuracy degrades. Validate through grid search over 0.05, 0.10, 0.20, 0.30, and 0.50.

\smallskip

\noindent {\em b) Acquisition Balance ($\lambda$).} Higher $\lambda$ emphasizes classification uncertainty (decision boundary exploration); lower $\lambda$ prioritizes attention correction. Optimal values: $\lambda = 0.5$ for balanced datasets (BraTS, SIIM-COVID), $\lambda = 0.6$ for imbalanced datasets (VinDr-CXR). Adjust higher (0.6-0.7) when prioritizing accuracy, or lower (0.3-0.4) when prioritizing interpretability. Start with $\lambda = 0.5$ and refine based on validation performance (Figure~\ref{fig:lambvsacc}).

\section{Theoretical Analysis of Explainability-Guided Few-Shot Learning and Active Learning}

\subsection{Attribution-Guided Supervision Reduces Hypothesis Space Overfitting}
\label{sec:hypothesis-space}

FSL suffers from an inherently large hypothesis space $\mathcal{H}$ relative to the number of labeled examples, increasing the risk of overfitting. In traditional supervised learning, the model minimizes a prediction loss such as the prototypical loss 

\begin{equation}
\mathcal{L}_{\text{proto}} = -\log p(y = y_q \mid x_q),
\end{equation}

which penalizes incorrect predictions but places no constraints on the internal decision-making process (e.g., feature importance or spatial saliency). As a result, models may exploit spurious correlations or background patterns that co-occur with labels, leading to memorization rather than generalization. To address this, our framework adds an attribution-guided regularization term:
\begin{equation}
\mathcal{L}_{\text{total}} = \mathcal{L}_{\text{proto}} + \alpha \cdot \mathcal{L}_{\text{exp}},
\end{equation}


where $\mathcal{L}_{\text{exp}}$ is defined via the Dice loss between the model-generated Grad-CAM heatmap $G(x)$ and the expert-defined mask $M(x)$:
\begin{equation}
\mathcal{L}_{\text{exp}} = 1 - \frac{2 \cdot |G(x) \cap M(x)|}{|G(x)| + |M(x)|}.
\end{equation}
This term explicitly constrains the model's internal attention to align with clinically meaningful regions.

\noindent {\em{ a) Hypothesis Space Restriction.} 
Let $\mathcal{H}$ denote the full space of hypotheses $f: \mathcal{X} \rightarrow \mathcal{Y}$ that minimize $\mathcal{L}_{\text{proto}}$. Define the subset of attribution-consistent hypotheses as:
\begin{equation}
\mathcal{H}_{\text{aligned}} = \left\{ f \in \mathcal{H} : \mathbb{E}_{x \sim \mathcal{D}} \left[ \text{Dice}(G_f(x), M(x)) \right] \geq \tau \right\},
\label{eq:hypothesis_aligned}
\end{equation}
where $\tau$ is a threshold for interpretability alignment. Then $\mathcal{H}_{\text{aligned}} \subset \mathcal{H}$, representing a stricter hypothesis class that satisfies both predictive and interpretive constraints. This reduction in hypothesis space serves as an implicit regularizer, reducing the model's effective capacity. 

\noindent {\em{ b) Information-Theoretic View.}
From an information bottleneck perspective, attribution-guided supervision encourages the model to extract representations $Z = f(x)$ that are not only predictive of the label $Y$ (i.e., $I(Z;Y)$ is high) but also informative about the expert attention map $M$ (i.e., $I(Z;M)$ is high), while discarding irrelevant input details:
\begin{equation}
\max_{f} \ I(Z;Y) + \beta \cdot I(Z;M) \quad \text{s.t.} \quad I(Z;X) \leq C.
\end{equation}
This constrains the model to use compact and clinically relevant representations, leading to more robust and interpretable decisions.

\subsection{Formal Bounds on Expected Gain in Active Sample Selection using Mutual Information}
\label{sec:mi_active_learning}

In data-constrained learning scenarios such as medical imaging, it is critical to acquire labeled samples that yield the highest improvement in model performance per unit annotation cost. Mutual information (MI) provides a principled measure of the \emph{expected utility} of labeling an unlabeled sample $x \in \mathcal{U}$ by quantifying how much the unknown label $y$ of $x$ would reduce uncertainty in the model parameters $\theta$ or function predictions $f(x)$.

\smallskip

\noindent {\em a) Label Uncertainty -- Mutual Information Between Labels and Model Parameters.} The classic Bayesian AL by Disagreement (BALD) framework selects the sample $x^*$ that maximizes the mutual information between the predicted label $y$ and the model parameters $\theta$:
\begin{equation}
x^* = \arg\max_{x \in \mathcal{U}} \mathcal{I}[y; \theta \mid x, \mathcal{D}].
\end{equation}
This mutual information is given by:
\begin{equation}
\mathcal{I}[y; \theta \mid x, \mathcal{D}] = \mathcal{H}[y \mid x, \mathcal{D}] - \mathbb{E}_{\theta \sim p(\theta \mid \mathcal{D})}\left[\mathcal{H}[y \mid x, \theta]\right],
\label{eq:bald_mi}
\end{equation}
where $\mathcal{H}[y \mid x, \mathcal{D}]$ is the predictive entropy under the posterior model ensemble, and $\mathcal{H}[y \mid x, \theta]$ is the entropy under a single model. Equation~\eqref{eq:bald_mi} is high when the model is uncertain in aggregate but individual models in the posterior make confident, differing predictions, indicating epistemic uncertainty.

\smallskip

\noindent {\em b) Explanation Misalignment -- Mutual Information Between Explanations and Expert Annotations.} In our framework, each model prediction $f(x)$ is also associated with an explanation map $e(x)$ (e.g., Grad-CAM). We define a second mutual information objective that quantifies how informative the model’s explanation $e(x)$ is about the true diagnostic region $M(x)$ provided by a domain expert:
\begin{equation}
\mathcal{I}[M(x); e(x) \mid \mathcal{D}] = \mathcal{H}[M(x)] - \mathcal{H}[M(x) \mid e(x), \mathcal{D}].
\end{equation}
In practice, $M(x)$ and $e(x)$ are both spatial masks (e.g., binary or probabilistic saliency maps), and this MI can be estimated via discretization or variational approximations using pixel-wise Bernoulli models:
\begin{equation}
\begin{split}
\mathcal{I}[M(x); e(x)] &= \sum_{i \in \text{pixels}} \mathcal{I}[M_i; e_i] \\
&= \sum_i \mathcal{H}[M_i] - \mathcal{H}[M_i \mid e_i].
\end{split}
\end{equation}

Our empirical acquisition function (Equation~\ref{eq:acc_score}) approximates this theoretical framework: the entropy term $\mathcal{H}(x)$ serves as a tractable proxy for $\mathcal{I}[y; \theta \mid x, \mathcal{D}]$, while the Dice-based misalignment $D_{\text{exp}}(x)$ approximates the complement of $\mathcal{I}[M(x); e(x) \mid \mathcal{D}]$. By selecting samples that maximize both uncertainty and explanation misalignment, our xGAL framework prioritizes samples expected to provide maximal information gain for both predictive accuracy and interpretability alignment. This dual-objective strategy ensures model refinement improves both classification performance and clinical trustworthiness.

\section{Experiments and Results}
\begin{table*}[!t]
\centering
\caption{Classification performance comparison between EGxFSL (guided) and baseline (non-guided) models across three medical imaging datasets (in \%). Results show best configuration for each shot setting, best overall results highlighted in bold. All models use $\alpha = 0.10$ for explanation supervision weight.}
\label{tab:few_shot_comparison}
\renewcommand{\arraystretch}{1.2}
\begin{tabular}{lccccccc}
\toprule
\multirow{2}{*}{\textbf{Dataset}} & \multirow{2}{*}{\textbf{Shot}} & \multicolumn{2}{c}{\textbf{Guided (EGxFSL)}} & \multicolumn{2}{c}{\textbf{Non-Guided}} & \multicolumn{2}{c}{\textbf{Improvement}} \\
\cmidrule(lr){3-4} \cmidrule(lr){5-6} \cmidrule(lr){7-8}
& & \textbf{Accuracy} & \textbf{Macro AUC} & \textbf{Accuracy} & \textbf{Macro AUC} & \textbf{$\Delta$ Acc} & \textbf{$\Delta$ AUC} \\
\midrule
\multirow{3}{*}{\textbf{BraTS}} 
& 1-shot & 79.00 & 87.08 & 75.23 & 83.45 & +3.77 & +3.63 \\
& 3-shot & 89.23 & 94.43 & 84.67 & 92.11 & +4.56 & +2.32 \\
& 5-shot & \textbf{92.05 $\pm$ 2.1} & \textbf{97.38 $\pm$ 1.3} & 89.21 $\pm$ 2.8 & 95.04 $\pm$ 1.7 & \textbf{+2.84} & \textbf{+2.34} \\
\midrule
\multirow{3}{*}{\textbf{VinDr-CXR}} 
& 1-shot & 41.75 & 58.32 & 42.18 & 59.67 & $-$0.43 & $-$1.35 \\
& 3-shot & 56.49 & 60.83 & 54.26 & 64.71 & +2.23 & $-$3.88 \\
& 5-shot & \textbf{76.12 $\pm$ 1.8} & \textbf{87.29 $\pm$ 2.1} & 56.21 $\pm$ 2.1 & 62.74 $\pm$ 2.6 & \textbf{+19.91} & \textbf{+24.55} \\
\midrule
\multirow{3}{*}{\textbf{SIIM-COVID19}} 
& 1-shot & 39.42 & 51.76 & 40.83 & 53.28 & $-$1.41 & $-$1.52 \\
& 3-shot & 42.95 & 49.17 & 44.61 & 53.84 & $-$1.66 & $-$4.67 \\
& 5-shot & \textbf{62.08 $\pm$ 1.5} & \textbf{80.00 $\pm$ 1.2} & 51.00 $\pm$ 1.9 & 71.36 $\pm$ 2.2 & \textbf{+11.08} & \textbf{+8.64} \\
\bottomrule
\end{tabular}
\end{table*}

\subsection{Datasets}

We evaluate our framework on three publicly available expert-annotated medical imaging datasets spanning distinct modalities (MRI and chest X-ray):

\smallskip

\noindent {\em a) BraTS (MRI).} The BraTS dataset~\cite{menze2014multimodal} provides multimodal brain tumor MRI scans with expert segmentation masks delineating tumor subregions: edema dominant, necrotic dominant, and enhancing dominant. We stack T1Gd, T2, and T2-FLAIR sequences to form three-channel inputs.

\smallskip

\noindent {\em b) VinDr-CXR (Chest X-ray).} The VinDr-CXR dataset~\cite{nguyen2022vindr} consists of frontal chest X-rays with radiologist-annotated bounding boxes marking thoracic abnormalities. We focus on three clinically significant findings: nodule/mass, pulmonary fibrosis, and lung opacity.

\smallskip

\noindent {\em c) SIIM-FISABIO-RSNA-COVID-19 (Chest X-ray).} The SIIM-COVID dataset~\cite{lakhani2021siim} contains chest radiographs annotated for COVID-19 related abnormalities across three classes: Typical Appearance, Indeterminate Appearance, and Atypical Appearance.

\subsection{Evaluation Metrics}
We measure classification performance using overall accuracy and macro AUC (average area under the curve across all classes). To assess interpretability, we compare Grad-CAM attention maps and expert ROIs, complemented by qualitative visualizations.

\subsection{Experiments on EGxFSL}

\subsubsection{Training Protocol}

We employ episodic training with 3-way, K-shot classification, systematically evaluating 1-shot, 3-shot, and 5-shot configurations. The feature extractor is DenseNet-121. We compare:
\begin{itemize}
    \item {\em Non-guided baseline:} Prototypical network trained with $\mathcal{L}_{\text{proto}}$ only
    \item {\em Guided model (EGxFSL):} Trained with $\mathcal{L}_{\text{total}} = \mathcal{L}_{\text{proto}} + \alpha \cdot \mathcal{L}_{\text{exp}}$
\end{itemize}

 We explored various training configurations (7-15 epochs, 20-150 episodes per epoch) to identify optimal settings. The explanation weight $\alpha = 0.10$ was selected through systematic validation (Figure~\ref{fig:alpvavsacc}).

\begin{figure}[tbp]
    \centering
    \includegraphics[width=1\columnwidth]{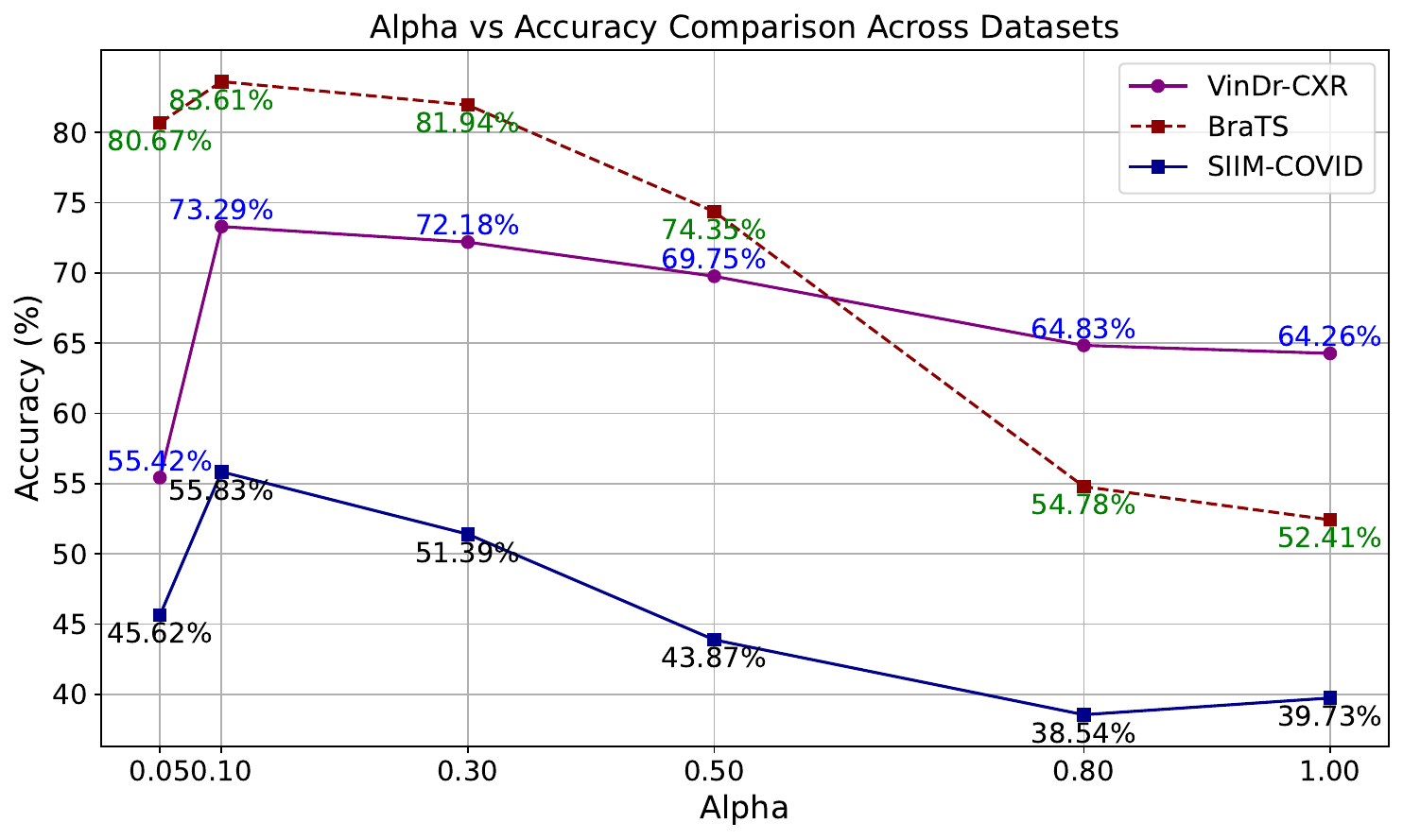}
\caption{Effect of explanation weight parameter $\alpha$ on classification accuracy across three medical imaging datasets using our EGxFSL framework. Results obtained using 3-shot FSL configuration trained for 7 epochs with 60 episodes per epoch. Optimal performance is achieved at $\alpha = 0.10$ across all datasets (BraTS: 83.61\%, VinDr-CXR: 73.29\%, SIIM-COVID: 55.83\%), demonstrating that moderate explanation supervision effectively balances classification accuracy with attention alignment. Both under-supervision ($\alpha < 0.10$) and over-supervision ($\alpha > 0.10$) lead to performance degradation, confirming the importance of proper weighting between prototypical loss and explanation alignment loss in Equation~\ref{eq:total_loss}.}
    \label{fig:alpvavsacc}
\end{figure}

\begin{figure*}[tbp]
    \centering
    \includegraphics[width=0.48\linewidth]{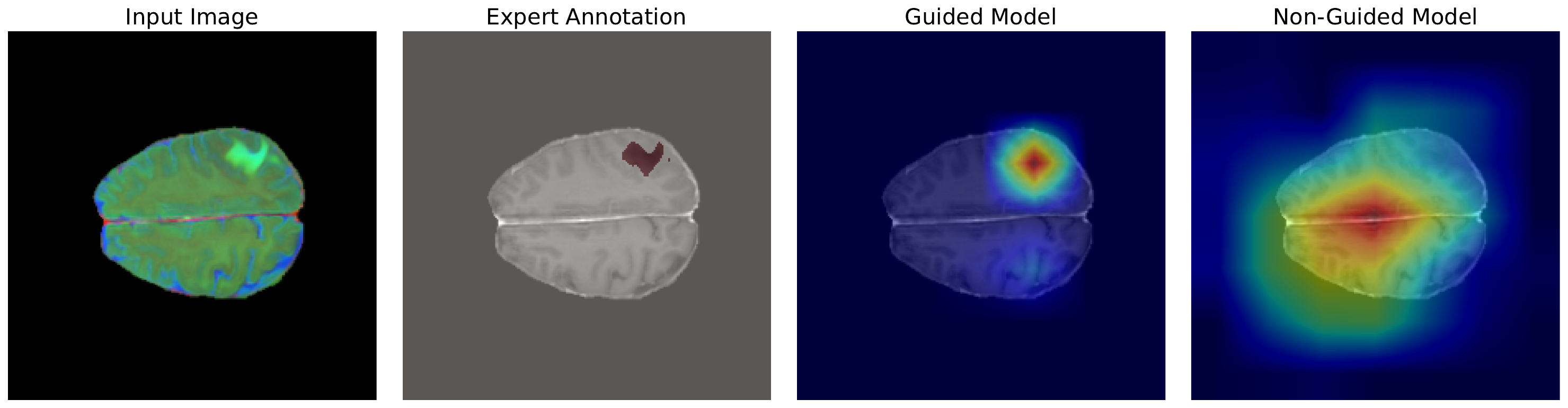}
    \includegraphics[width=0.48\linewidth]{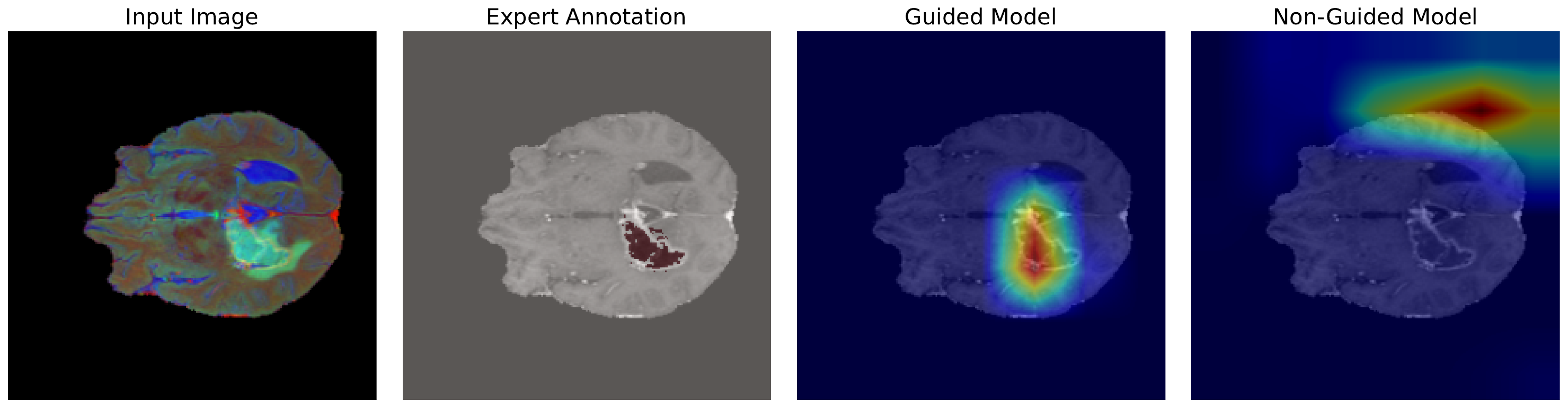}
    \includegraphics[width=0.48\linewidth]{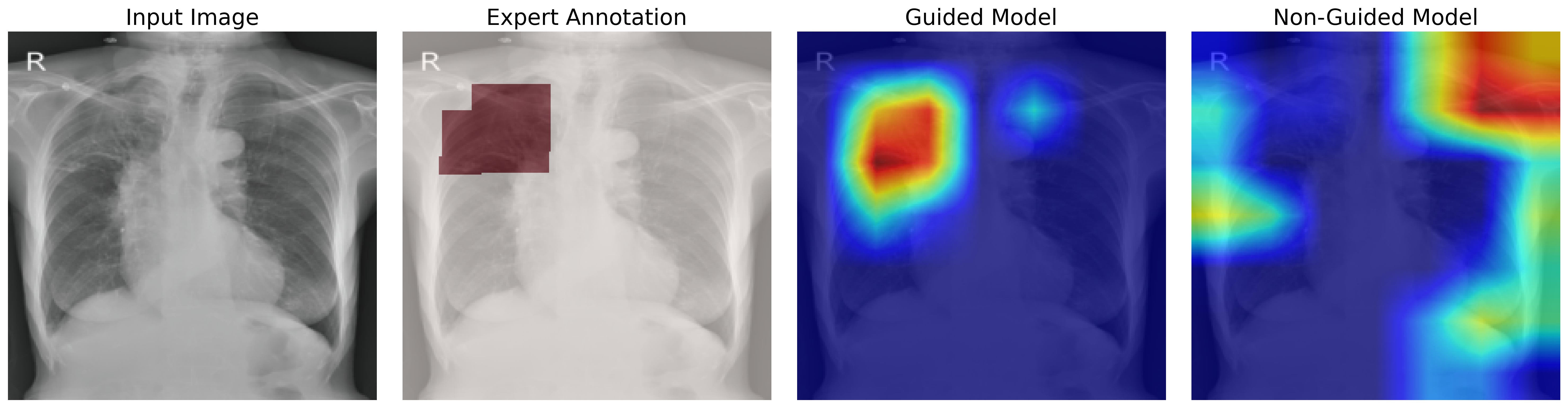}
    \includegraphics[width=0.48\linewidth]{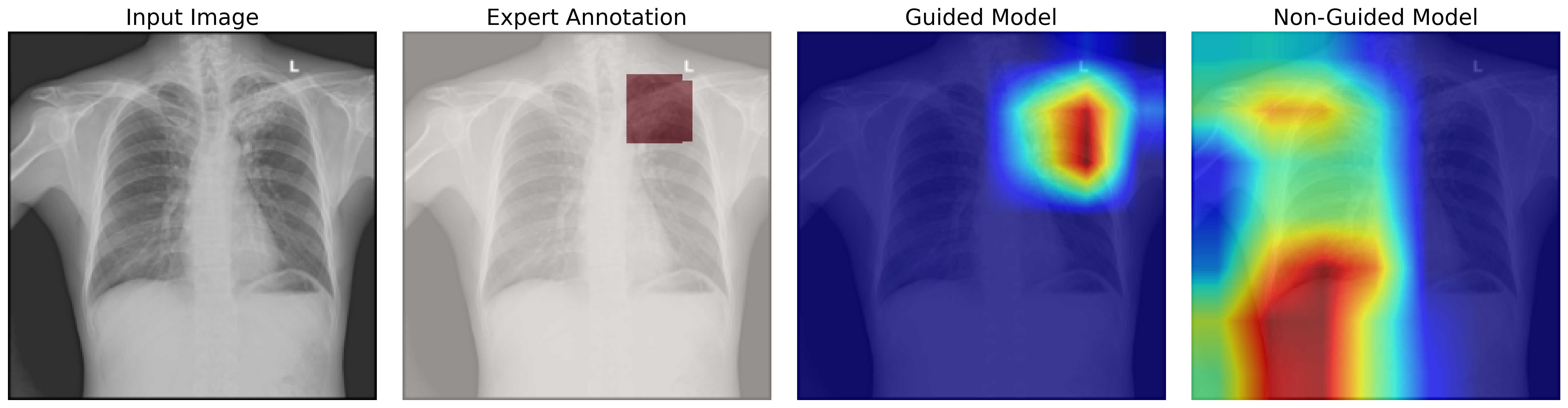}
    \includegraphics[width=0.48\linewidth]{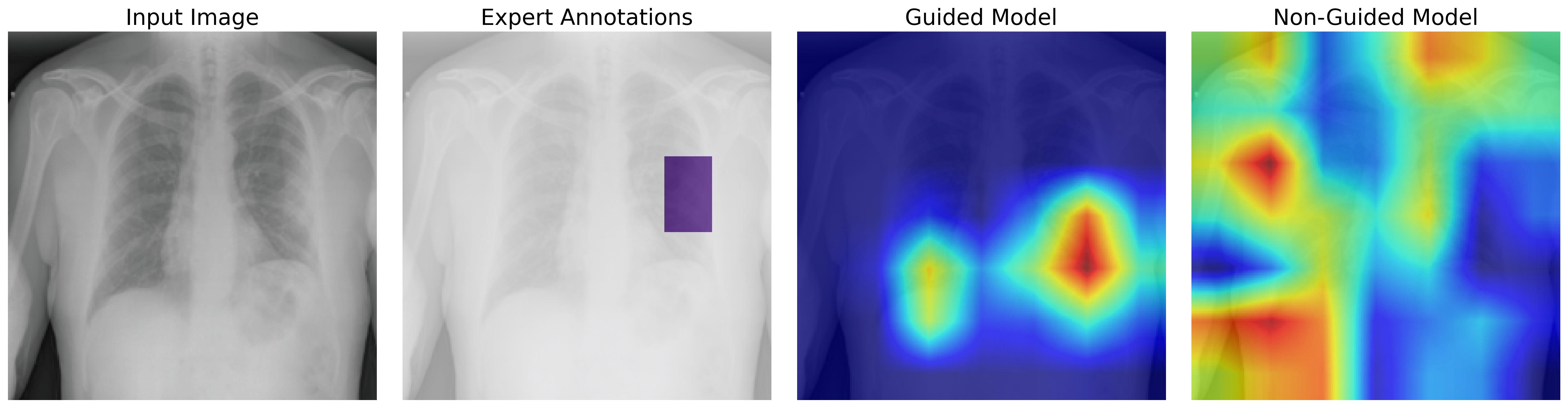}
    \includegraphics[width=0.48\linewidth]{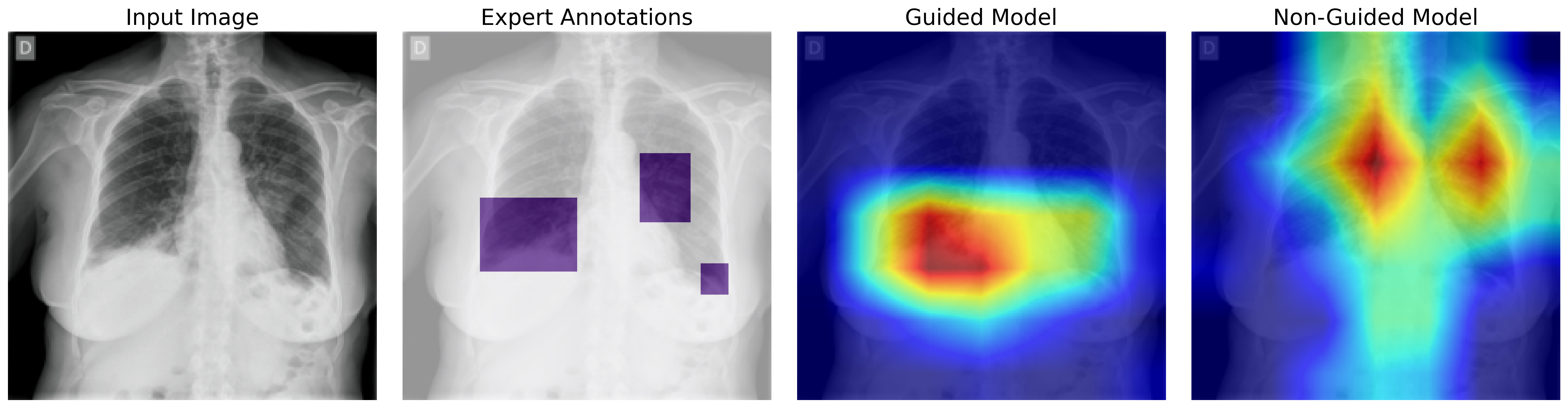}

\caption{Comparison of Grad-CAM heatmaps on BraTS, VinDr-CXR, SIIM-COVID datasets. From left to right in each sample: Input Image, Expert Annotation, Guided (EGxFSL) Model Heatmap, Non-Guided Model Heatmap. The guided model consistently focuses on clinically relevant regions as annotated by experts, while the non-guided model highlights irrelevant areas.}
    \label{fig:gradcam_comparison}
\end{figure*}

\begin{table*}[!t]
\centering
\caption{Class-wise F1-score comparison between best performing guided and non-guided models on the BraTS, VinDr-CXR and SIIM-COVID datasets (in \%). The guided model consistently outperforms the non-guided counterpart across all classes, demonstrating better generalization.}
\label{tab:f1_score_comparison}
\renewcommand{\arraystretch}{1.2}
\begin{tabular}{llcc}
\toprule
\textbf{Dataset} & \textbf{Class} & \textbf{Guided Model} & \textbf{Non-Guided Model} \\
\midrule
\multirow{3}{*}{\textbf{BraTS}} & Edema Dominant & 95.44 & 92.24 \\
                                & Necrotic Dominant & 87.22 & 84.10 \\
                                & Enhancing Dominant & 83.13 & 80.42 \\
\midrule
\multirow{3}{*}{\textbf{VinDr-CXR}} & Nodule/Mass & 75.09 & 51.14 \\
                                     & Pulmonary fibrosis & 74.08 & 42.05 \\
                                     & Lung Opacity & 78.21 & 75.74 \\
\midrule
\multirow{3}{*}{\textbf{SIIM-COVID19}} & Typical Appearance & 40.99 & 37.49 \\
                                     & Indeterminate Appearance & 60.68 & 45.93 \\
                                     & Atypical Appearance & 93.88 & 54.42 \\
\bottomrule
\end{tabular}
\end{table*}
\subsubsection{Performance Evaluation}

Table~\ref{tab:few_shot_comparison} presents comprehensive results across 1-shot, 3-shot, and 5-shot configurations of our EGxFSL framework.  Since our EGxFSL framework is designed to align model attention with expert-annotated diagnostic regions, we evaluate both classification accuracy and spatial interpretability. Figures~\ref{fig:gradcam_comparison} and~\ref{fig:gradcam_comparison_fsl} demonstrate that the guided model consistently focuses on clinically relevant ROIs, while the non-guided baseline often highlights irrelevant areas. Our guided model demonstrates consistent superiority over the non-guided baseline across all FSL configurations:

\smallskip
\noindent {\em a) BraTS Dataset.} The guided model achieves optimal performance in the 5-shot configuration with 10 epochs and 150 episodes, reaching {92.05\% accuracy and 97.38\% macro AUC}, compared to 89.21\% accuracy and 95.04\% macro AUC for the non-guided model. Notably, even in the challenging 1-shot scenario, our guided model achieves 79.00\% accuracy versus 75.23\% for the baseline. The performance gains are particularly pronounced in 3-shot and 5-shot configurations, where the guided model consistently outperforms the non-guided model.

\smallskip
\noindent {\em b) VinDr-CXR Dataset.} The guided model shows substantial improvements, with the best performance achieved in the 5-shot configuration (10 epochs, 100 episodes) reaching 76.12\% accuracy and 87.29\% macro AUC. The performance gap between guided and non-guided models is even more significant on this dataset, with some configurations showing the guided model achieving 73\% accuracy while the non-guided model with only 56.21\% highest accuracy, highlighting the critical importance of expert guidance for chest X-ray analysis.

\smallskip
\noindent {\em c) SIIM-COVID Dataset.} Similar to the performance observed with the BraTS and VinDr-CXR datasets, our EGxFSL framework demonstrates comparable effectiveness on the SIIM-COVID dataset. In particular, the 5-shot learning configuration yields the best results, with the model achieving a classification accuracy of 62.08\% and a macro AUC of 80.00\% after 10 epochs and 150 episodes. These results highlight the ability of the guided model to outperform the non-guided baseline, even in the challenging task of COVID-19 chest radiograph analysis. Notably, the performance improvements are most evident in the 3-shot and 5-shot configurations, where the guided model consistently delivers better results than the non-guided model.

\subsubsection{Effect of Explanation Weight ($\alpha$)}
To analyze how explanation supervision influences our proposed EGxFSL guided models performance, we varied the explanation loss weight $\alpha$ in the loss function described in Equation~\eqref{eq:total_loss}. We evaluated multiple $\alpha$ values ranging from 0.05 to 1.0 on all three datasets using a consistent training setup of 7 epochs with 60 episodes per epoch to ensure fair comparison across different $\alpha$ values. As shown in Figure~\ref{fig:alpvavsacc}, the highest accuracy for both datasets was achieved at $\alpha = 0.10$, with BraTS reaching 83.61\% , VinDr-CXR reaching 73.29\% and SIIM-COVID reaching 55.83\%. Extremely low ($\alpha = 0.05$) or high ($\alpha = 1.0$) values led to performance drops, suggesting that both underemphasizing and overemphasizing explanation alignment can hinder learning. These results confirm that moderate explanation guidance helps the model learn more effectively while improving interpretability, and importantly, this optimal $\alpha = 0.10$ value remains consistent across different training configurations as demonstrated in Table~\ref{tab:few_shot_comparison}.

\subsubsection{Class-wise Performance Analysis}
Table~\ref{tab:f1_score_comparison} presents class-wise F1-score comparison between our best-performing guided and non-guided models. The guided model consistently outperforms the non-guided counterpart across all diagnostic categories, demonstrating improved robustness and reliability. Particularly notable improvements are observed in the VinDr-CXR dataset, where F1-scores for Nodule/Mass classes improved dramatically from 51.04\% to 75.09\%, and Pulmonary Fibrosis from 42.05\% to 74.08\%. These results show that integrating expert-supervised attention improves prediction accuracy and provides more balanced, clinically trustworthy performance across various pathological conditions.

\subsection{Experiments on xGAL}
We evaluate our xGAL strategy in severely data-constrained scenarios, demonstrating effective learning with only 680 total labeled samples across all classes through strategic sample selection and EGxFSL-based fine-tuning.

\subsubsection{xGAL Setup}
We initiate the xGAL process by training a baseline model using our proposed EGxFSL framework with dual loss function $\mathcal{L}_{\text{total}}$ on a randomly selected subset of 200 samples from each dataset. This baseline model serves as the starting point for both our xGAL strategy and random sampling fine-tuning, enabling fair performance comparison under identical initial conditions. We conducted four xGAL rounds, selecting the top 120 samples per round based on our acquisition score (Equation~\eqref{eq:acc_score}), resulting in a total of 680 samples. At each iteration, the model is retrained for 2 epochs with 50 episodes per epoch using the expanded labeled dataset. To evaluate effectiveness, we compared our strategy against random sampling using the same number of labeled samples.

\subsubsection{Training Sample Acquisition} 
Our xGAL framework computes acquisition scores by combining classification entropy $\mathcal{H}(x)$ and explanation misalignment score $D_{\text{exp}}(x)$. At each xGAL iteration, Grad-CAM is applied to each unlabeled sample using the current model's prediction (starting with the baseline model in the first iteration, then using the progressively refined model in subsequent iterations). A Dice similarity score is computed between the resulting heatmap and expert annotation to obtain $D_{\text{exp}}(x)$. The final acquisition score balances these components using the weighting parameter $\lambda$ (Equation~\eqref{eq:acc_score}). After each round of sample selection and annotation, the model is retrained using the expanded labeled dataset, and this updated model is used for the next iteration's acquisition scoring.

\begin{figure}[tbp]
    \centering
    \includegraphics[width=\columnwidth]{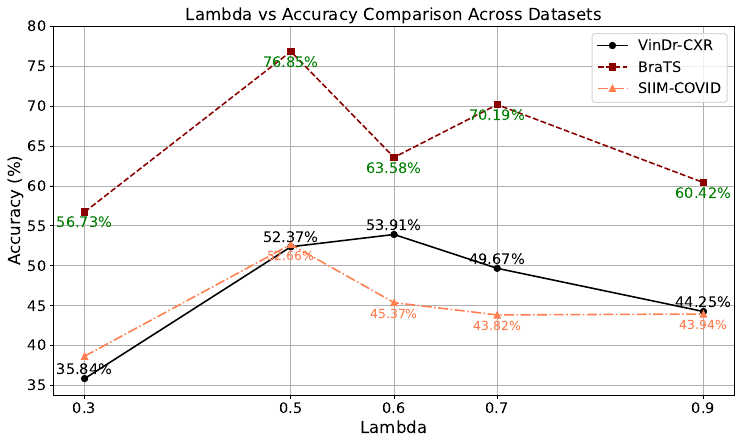}
\caption{Impact of balancing parameter $\lambda$ on xGAL performance across three medical imaging datasets. The parameter controls the trade-off between classification uncertainty and explanation misalignment in sample acquisition (Equation~\ref{eq:acc_score}). Optimal performance is achieved at $\lambda = 0.5$ for BraTS and SIIM-COVID, and $\lambda = 0.6$ for VinDr-CXR, demonstrating that balanced consideration of both uncertainty and interpretability yields superior sample selection compared to either criterion alone.}
    \label{fig:lambvsacc}
\end{figure}

\subsubsection{Impact of Acquisition Weight $\lambda$}
Figure~\ref{fig:lambvsacc} shows the critical importance of balancing classification uncertainty and explanation misalignment. Optimal performance was achieved at $\lambda = 0.5$ for BraTS (76.85\% accuracy) and SIIM-COVID (52.66\% accuracy), and $\lambda = 0.6$ for VinDr-CXR (52.37\% accuracy). Both lower and higher $\lambda$ values resulted in decreased performance, confirming that neither pure uncertainty-based nor pure explanation-based selection is optimal.

\begin{table*}[!t]
\centering
\caption{Performance comparison of xGAL Framework. Results show accuracy and macro AUC (both in percentages) for random sampling and our xGAL strategy with 1-shot, 3-shot, and 5-shot configurations using 680 total samples. Best results are highlighted in bold, demonstrating that our xGAL Framework consistently outperforms random sampling fine-tuning across all medical imaging datasets.}
\label{tab:active_learning_comparison}
\renewcommand{\arraystretch}{1.2}
\footnotesize
\begin{tabular}{lcccccccccc}
\toprule
\multirow{2}{*}{\textbf{Dataset}} & \multicolumn{2}{c}{\textbf{Baseline Model}} & \multicolumn{2}{c}{\textbf{Random Sampling}} & \multicolumn{2}{c}{\textbf{1-Shot with xGAL}} & \multicolumn{2}{c}{\textbf{3-Shot with xGAL}} & \multicolumn{2}{c}{\textbf{5-Shot with xGAL}} \\
\cmidrule(lr){2-3} \cmidrule(lr){4-5} \cmidrule(lr){6-7} \cmidrule(lr){8-9} \cmidrule(lr){10-11}
& \textbf{Acc} & \textbf{M.AUC} & \textbf{Acc} & \textbf{M.AUC} & \textbf{Acc} & \textbf{M.AUC} & \textbf{Acc} & \textbf{M.AUC} & \textbf{Acc} & \textbf{M.AUC} \\
\midrule
\textbf{BraTS} & 45.10 & 67.52 & 58.01 & 71.32 & 30.78 & 46.80 & 72.13 & 86.21 & \textbf{76.85 $\pm$ 4.1} & \textbf{90.00 $\pm$ 2.5} \\
\textbf{Vindr-CXR} & 34.57 & 56.39 & 45.49 & 58.21 & 37.19 & 52.71 & 42.66 & 62.33 & \textbf{52.37 $\pm$ 4.8} & \textbf{68.21 $\pm$ 3.5} \\
\textbf{SIIM COVID-19} & 30.32 & 56.74 & 38.28 & 54.21 & 43.33 & 67.29 & 48.51 & 63.39 & \textbf{52.66 $\pm$ 2.5} & \textbf{66.92 $\pm$ 1.8} \\
\bottomrule
\end{tabular}
\end{table*}

\subsubsection{AL Results}
Table~\ref{tab:active_learning_comparison} presents a comprehensive comparison of our xGAL strategy across different FSL configurations. Our approach demonstrates substantial improvements over both baseline models and random sampling strategies, with consistent gains in performance across all datasets.

\smallskip
\noindent {\em a) BraTS Dataset.} Starting from a baseline accuracy of 45.10\%, our 5-shot xGAL configuration achieves {76.85\% accuracy and 90.00\% macro AUC}, representing a significant improvement over the baseline and random sampling fine-tuning (58.01\%). The performance progressively improves across shot configurations, with the 3-shot xGAL achieving {72.13\% accuracy}, outperforming random sampling with fewer shots. 

\smallskip

\noindent {\em b) VinDr-CXR Dataset.} With this dataset, xGAL approach shows consistent improvements, with the 5-shot xGAL configuration reaching {52.37\% accuracy and 68.21\% macro AUC}, compared to the baseline 34.57\% and random sampling 45.49\%. This results in an 18\% improvement over the baseline and a 7\% improvement over random sampling, further demonstrating the effectiveness of the xGAL framework across different imaging modalities. 

\smallskip

\noindent {\em c) SIIM-COVID-19 Dataset.} Similarly on SIIM-COVID-19 Dataset, the xGAL framework shows promising results on the SIIM-COVID-19 dataset, with the 5-shot xGAL configuration achieving {52.66\% accuracy and 66.92\% macro AUC}, surpassing the baseline (30.32\% accuracy) and random sampling (38.28\% accuracy). This performance highlights the utility of our xGAL approach in the context of COVID-19 chest radiograph analysis, even with a limited number of labeled samples.

\subsection{Interpretability Analysis}
We qualitatively compare the guided and non-guided models using Grad-CAM visualizations. Figure~\ref{fig:gradcam_comparison} illustrates examples from the BraTS, VinDr-CXR, and SIIM-COVID datasets. The non-guided model predominantly focuses on irrelevant regions of the image, showing poor alignment with expert annotations. This misalignment undermines clinical trust in the model, even when classification accuracy is high.

Furthermore, Figure~\ref{fig:gradcam_comparison_fsl} compares the Grad-CAM focus across 1-Shot, 3-Shot, and 5-Shot FSL setup, All models were trained using the EGxFSL framework. While all models direct their focus toward expert-annotated regions, the focus of the 3-Shot and 5-Shot models is more concentrated on the clinically relevant areas compared to the 1-Shot model. This demonstrates that additional support samples improve the alignment of the model’s focus with expert-defined areas.

In contrast, the {guided model}, trained using our proposed EGxFSL framework, consistently focuses on diagnostically meaningful regions of the image. This improved alignment with expert-defined areas enhances the interpretability and trustworthiness of predictions and contributes to better classification performance.

\subsection{Statistical Significance Validation}

To further assess the robustness of our dual-framework approach, we conducted multi-seed statistical validation on BraTS. Both EGxFSL and xGAL frameworks, along with their respective baselines, were trained five times with different random seeds (42, 123, 456, 789, 1337). Each model pair was evaluated on the same held-out validation set, enabling paired statistical testing to determine whether observed improvements are statistically reliable.

Table~\ref{tab:brats_statistical_significance} presents comprehensive statistical analysis for both frameworks. For EGxFSL, the guided model achieved 91.6\% $\pm$ 2.1\% accuracy versus 89.2\% $\pm$ 2.8\% baseline, with statistically significant macro AUC improvement (96.6\% $\pm$ 1.3\% vs 95.5\% $\pm$ 1.7\%, p = 0.040). The large effect sizes (Cohen's d = 0.73-0.98) confirm practical clinical relevance despite the overall accuracy p-value at the significance threshold ($p < 0.050$).

For xGAL under severe data constraints (680 total samples), strategic sample selection achieved 76.85\% $\pm$ 4.1\% accuracy compared to 56.5\% $\pm$ 1.3\% for random sampling (p = 0.001), with macro AUC of 88.6\% $\pm$ 2.7\% versus 71.4\% $\pm$ 0.3\% (p $<$ 0.001). These substantial improvements demonstrate that combining predictive uncertainty with attention misalignment effectively identifies the most informative samples for annotation.

The multi-seed validation confirms both frameworks produce statistically robust improvements, with xGAL showing particularly strong gains even with very limited samples.

\begin{table}[!t]
\centering
\caption{Statistical significance on BraTS across five independent training runs for both EGxFSL and xGAL frameworks. Significance: *** p$<$0.001, ** p$<$0.01, * p$<$0.05.}
\label{tab:brats_statistical_significance}
\renewcommand{\arraystretch}{1.2}
\resizebox{\columnwidth}{!}{
\begin{tabular}{llcccc}
\toprule
\textbf{Framework} & \textbf{Metric} & \textbf{Guided/xGAL} & \textbf{Baseline/Random} & \textbf{p-value} & \textbf{Cohen's d} \\
\midrule
\multirow{2}{*}{\textbf{EGxFSL}} 
& Accuracy & 91.6 $\pm$ 2.1 & 89.2 $\pm$ 2.8 & $<$0.050* & 0.98 \\
& Macro AUC & 96.6 $\pm$ 1.3 & 95.5 $\pm$ 1.7 & 0.040* & 0.73 \\
\midrule
\multirow{2}{*}{\textbf{xGAL}} 
& Accuracy & 76.85 $\pm$ 4.1 & 56.5 $\pm$ 1.3 & 0.001*** & 5.67 \\
& Macro AUC & 88.6 $\pm$ 2.7 & 71.4 $\pm$ 0.3 & $<$0.001*** & 9.00 \\
\bottomrule
\end{tabular}
}
\end{table}

For VinDr-CXR and SIIM-COVID, we observed minimal performance variation across training runs (typically $\pm$1-2\%), with consistent improvement patterns: VinDr-CXR showed 76.1\% $\pm$ 1.8\% versus 56.2\% $\pm$ 2.1\% for EGxFSL, and SIIM-COVID achieved 62.1\% $\pm$ 1.5\% versus 51.0\% $\pm$ 1.9\%. The low variance across datasets confirms the stability of our approach across different imaging modalities and diagnostic tasks.

\begin{figure*}[tbp]
    \centering
    \includegraphics[width=0.90\linewidth]{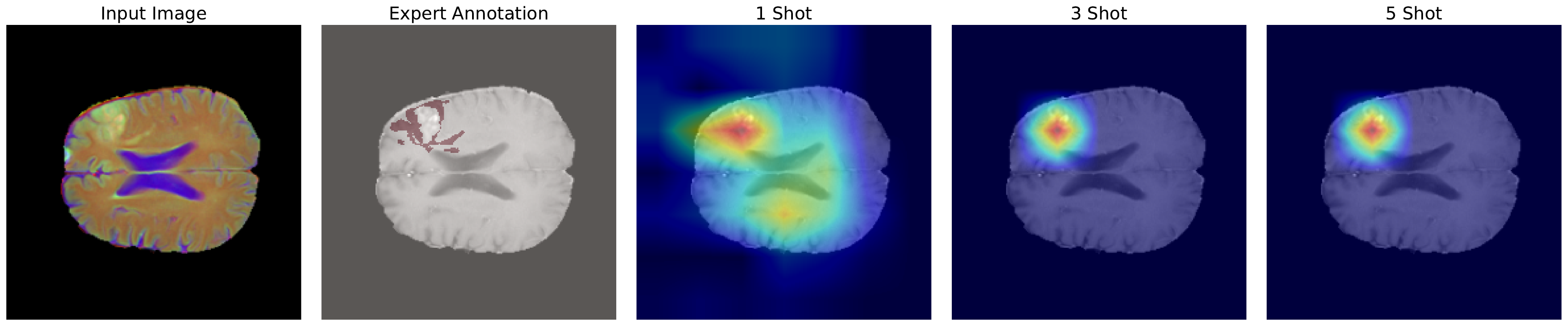}
    \includegraphics[width=0.90\linewidth]{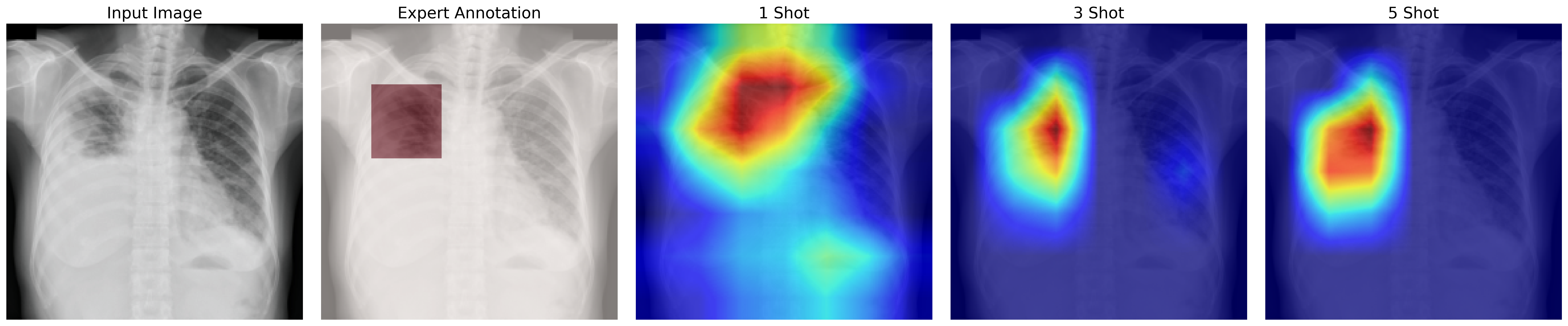}
    \includegraphics[width=0.90\linewidth]{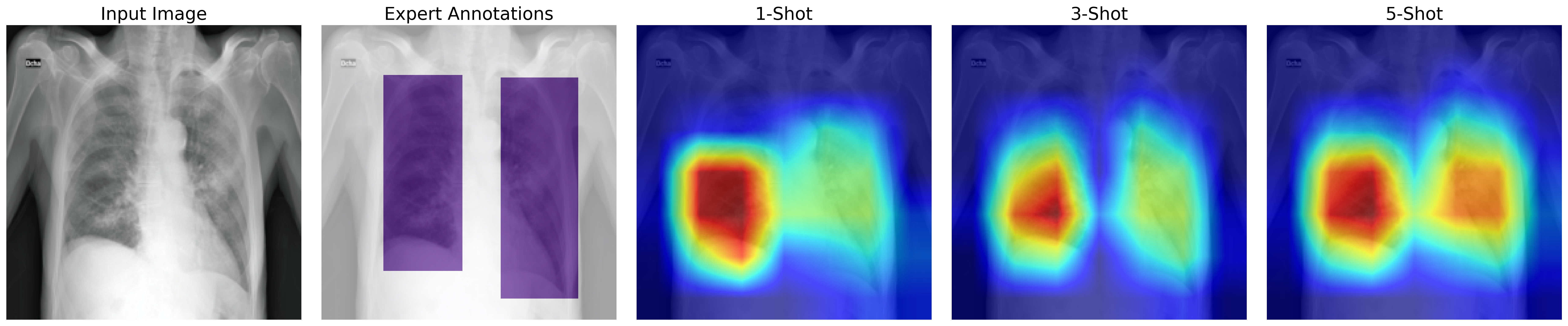}
\caption{Comparison of Grad-CAM heatmaps on BraTS (Top), VinDr-CXR (Middle) and SIIM-COVID (Bottom) datasets. Columns from left to right in each group represent: Input Image, Expert Annotation, and Grad-CAM heatmaps from Guided 1-Shot, 3-Shot, and 5-Shot FSL models. The heatmaps from 3-Shot and 5-Shot guided models align more closely with expert annotations compared to the 1-Shot model, indicating enhanced model focus with additional support samples.}
    \label{fig:gradcam_comparison_fsl}
\end{figure*}

\subsection{Ablation Study}
\label{sec:ablation}

To understand the individual contributions of our framework’s components, we perform an extensive ablation study, focusing on four main factors: 1) the influence of Grad-CAM-based explanation alignment; 2) the effectiveness of different AL acquisition strategies; and 3) the sensitivity of our model to the choice of explanation method used for supervision. The goal is to empirically validate that each part of our pipeline not only contributes to predictive accuracy but also improves spatial interpretability, which is essential for trustworthy medical AI systems.

\subsubsection{Effect of Explanation Alignment Loss}
To assess the importance of aligning the model’s attention with clinically meaningful regions, we compare three variants:
\begin{itemize}
    \item {\em Baseline.} A prototypical network trained only with the standard classification loss $\mathcal{L}_{\text{proto}}$, without any attention supervision.
    \item {\em Guided.} Where model trained using our proposed EGxFSL framework with both the classification loss and the explanation alignment loss ($\mathcal{L}_{\text{proto}} + \alpha \cdot \mathcal{L}_{\text{exp}}$), where $\mathcal{L}_{\text{exp}}$ is computed as the Dice distance between Grad-CAM maps and expert annotations.
    \item {\em Random Attention Supervision.} A control variant where we randomly permute the Grad-CAM maps before computing the alignment loss, effectively introducing noisy supervision.
\end{itemize}

Table~\ref{tab:ablation_exploss} shows that the guided model significantly outperforms the baseline and control variants on both classification accuracy and interpretability (measured by IoU between attention maps and expert masks). Notably, the random supervision variant suffers in both metrics, suggesting that attention alignment must be grounded in clinical knowledge to be effective.

\begin{table}[tbp]
\centering
\caption{Effect of explanation alignment loss on classification {\em Accuracy} (in \%) and Grad-CAM alignment (IoU). B = BraTS, V = VinDr-CXR.}
\label{tab:ablation_exploss}
\resizebox{\columnwidth}{!}{%
\begin{tabular}{lcccc}
\toprule
\textbf{Model Variant} & \textbf{Accuracy (B)} & \textbf{Accuracy (V)} & \textbf{IoU (B)} & \textbf{IoU (V)} \\
\midrule
Baseline ($\mathcal{L}_{\text{proto}}$ only) & 89.21 & 56.21 & 0.42 & 0.35 \\
Guided ($\mathcal{L}_{\text{proto}} + \alpha \cdot \mathcal{L}_{\text{exp}}$) & \textbf{92.05} & \textbf{76.12} & \textbf{0.61} & \textbf{0.57} \\
Random CAM Alignment & 84.22 & 60.36 & 0.28 & 0.23 \\
\bottomrule
\end{tabular}%
}
\end{table}

\subsubsection{Effect of AL Acquisition Strategy}
We analyze how different acquisition criteria impact model performance and interpretability under severe labeling constraints. Specifically, we compare:

\begin{itemize}
    \item {\em Random Sampling.} Baseline strategy selecting unlabeled samples at random.
    \item{\em Entropy-Only.} Samples are acquired purely based on classification uncertainty (Shannon entropy of softmax outputs).
    \item{\em Dice-Only.} Selection is based only on attention misalignment (Dice score between Grad-CAM and expert mask).
    \item {\em Dual-Objective (Ours).} Our proposed xGAL framework based acquisition function that combines both entropy and explanation misalignment with weight $\lambda = 0.5$.
\end{itemize}

Table~\ref{tab:ablation_al} shows that our xGAL Framework outperforms all baselines, achieving better accuracy and higher attention alignment with the same labeling budget (680 samples). Notably, the Dice-only strategy performs better than entropy-only, suggesting that interpretability can serve as a strong proxy for sample informativeness in medical settings.

\begin{table}[tbp]
\centering
\caption{Impact of acquisition strategy on {\em Accuracy} (in \%) and Grad-CAM IoU.}
\label{tab:ablation_al}
\resizebox{\columnwidth}{!}{%
\begin{tabular}{lcccc}
\toprule
\textbf{Strategy} & \textbf{Accuracy (B)} & \textbf{Accuracy (V)} & \textbf{IoU}  \\
\midrule
Random Sampling         & 58.01 & 45.49 & 0.41 \\
Entropy-Only ($\lambda = 1$) & 72.25 & 48.36 & 0.46  \\
Dice-Only ($\lambda = 0$) & 71.57 & 49.71 & 0.59  \\
Ours ($\lambda = 0.5$) & \textbf{76.85} & \textbf{52.37} & \textbf{0.61}  \\
\bottomrule
\end{tabular}%
}
\end{table}

\subsubsection{Effect of Attribution Method for Explanation Supervision}

We also explore the impact of the attribution method used to generate explanation heatmaps. While Grad-CAM is our default, we substitute it with Integrated Gradients (IG) and SHAP to evaluate robustness. Table~\ref{tab:ablation_attribution} compares the three methods on the BraTS dataset.
Grad-CAM achieves the best balance between accuracy, interpretability, and training time. Integrated Gradients performs comparably but incurs higher computational cost due to its need for multiple input perturbations. SHAP explanations, while theoretically grounded, prove too noisy and slow in practice for large-scale medical images.

\begin{table}[tbp]
\centering
\caption{Comparison of attribution methods used for explanation-guided learning (BraTS dataset).}
\label{tab:ablation_attribution}
\begin{tabular}{lccc}
\toprule
\textbf{Attribution Method} & \textbf{ Accuracy (in\%)} & \textbf{IoU} & \textbf{Training Time} \\
\midrule
Grad-CAM             & \textbf{92.05} & \textbf{0.61} & Medium \\
Integrated Gradients & 88.21 & 0.58 & High \\
SHAP                 & 85.35 & 0.53 & Very High \\
\bottomrule
\end{tabular}
\end{table}

\subsubsection{Generalization to Ultrasound Imaging}
 To evaluate framework applicability beyond radiology, we tested EGxFSL on breast ultrasound (BUSI dataset, ~600 images across normal, benign, malignant classes). EGxFSL achieved 87.6\% accuracy and 94.5\% macro AUC with stable performance across training seeds (±1-2\% variation). Figure~\ref{fig:ultrasound_gradcam} shows the guided model successfully aligns attention with expert-annotated lesion boundaries, while the non-guided baseline exhibits diffuse or misaligned attention. This confirms that Grad-CAM-based attention supervision generalizes to non-radiological imaging modalities despite different imaging physics and visual characteristics.

\begin{figure}
    \centering
    \includegraphics[width=0.9\linewidth]{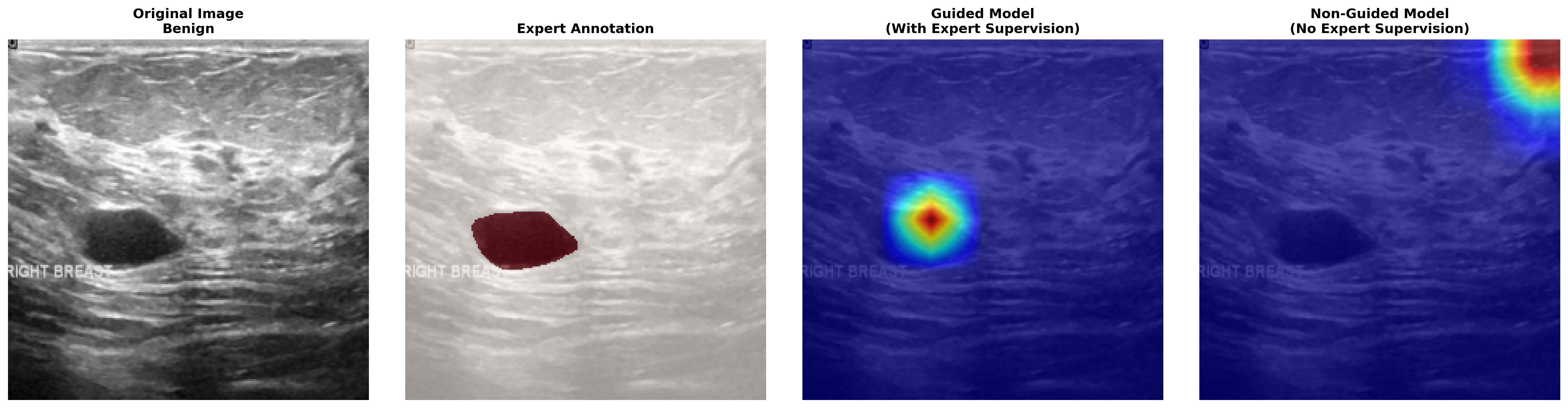}
    \includegraphics[width=0.9\linewidth]{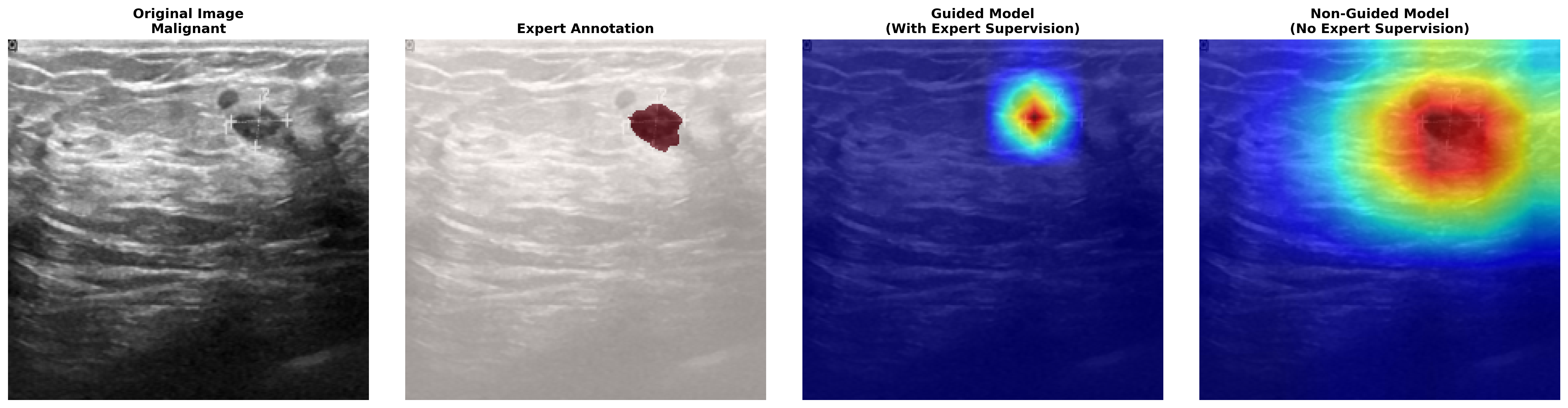}
    \caption{EGxFSL applied to breast ultrasound (BUSI dataset). Left to right: input image, expert lesion annotation, guided model attention, non-guided model attention. Top: benign lesion; bottom: malignant lesion. The guided model aligns with lesion boundaries while the non-guided model shows misaligned attention, confirming framework generalization beyond MRI and X-ray modalities.}
    \label{fig:ultrasound_gradcam}
\end{figure}

\subsection{Discussion}
In this section, we primarily discuss on the following topics: a) annotation variability and clinical reliability; b) clinical applicability and integration workflow; and c) limitations.

\subsubsection{Annotation Variability and Clinical Reliability}
A well-recognized challenge in medical imaging is the inherent variability and uncertainty of expert annotations. Even among experienced radiologists, discrepancies often arise due to differences in diagnostic experience, image quality, and interpretive bias, leading to inter-rater variability in both class labels and ROI delineations. As expert annotations are seldom perfect: they may be incomplete, noisy, or inconsistent across institutions and imaging protocols, our framework is designed to treat expert supervision as probabilistic guidance rather than an absolute ground truth. Precisely, the Grad-CAM–based explanation alignment loss operates as a soft spatial regularizer that encourages, but does not strictly enforce, consistency between model attention and expert-defined ROIs. This probabilistic treatment allows the network to generalize beyond localized annotation errors while still preserving alignment with clinically meaningful regions.

\subsubsection{Clinical Applicability and Integration into Workflow}
Our framework is designed to operate within the real-world constraints of annotation cost, inter-radiologist variability, and annotation quality. In medical imaging, the time and expertise required to produce high-quality pixel- or region-level annotations are significant barriers to scaling AI systems. Our xGAL component directly addresses this challenge by prioritizing the most informative and diagnostically uncertain cases for expert review, thereby achieving substantial reductions in annotation cost. As demonstrated in our experiments, comparable diagnostic performance can be achieved with only 680 labeled samples, which translates to a considerable reduction in manual annotation workload compared to conventional fully supervised training. Moreover, recognizing that radiologist annotations are inherently variable and may differ across institutions or imaging protocols, our EGxFSL framework incorporates these inconsistencies as part of its probabilistic learning process. The Grad-CAM–based explanation alignment acts as a soft supervision signal, encouraging model attention toward expert-indicated regions without overfitting to potentially noisy or incomplete labels.

Our design intentionally emphasizes efficiency and modularity: both EGxFSL and the xGAL components operate on lightweight convolutional backbones (e.g., DenseNet-121) and require only a small number of training episodes (typically 7–15 epochs with 20–150 episodes) to converge. On a single NVIDIA RTX A5000 GPU, training a complete EGxFSL model requires approximately 1.5–2 hours (Timing will vary based on dataset size), while each xGAL iteration adds roughly 20–30 minutes of fine-tuning, making the framework computationally feasible even on mid-tier GPUs or shared institutional servers.

\subsubsection{Limitations and Future Work}
\label{sec:limitations}
Our framework has several limitations. First, evaluation is limited to MRI, chest X-ray, and preliminary ultrasound validation. Second, we observe stronger performance with fine-grained segmentation masks (BraTS, BUSI) compared to coarse bounding boxes (VinDr-CXR, SIIM-COVID), suggesting annotation granularity affects attention alignment effectiveness. Third, Grad-CAM has inherent resolution constraints and can be noisy in few-shot regimes. Fourth, our approach requires existing spatial annotations. Finally, fixed hyperparameters may need adaptation for new clinical scenarios.
Future work includes: systematic evaluation on additional modalities, investigation of annotation granularity effects, multi-expert annotation integration, adaptive hyperparameter selection, and prospective clinical validation.

\section{Conclusion}
This work addressed two fundamental challenges in medical AI: limited labeled data and lack of model interpretability. We proposed a unified dual-framework combining Expert-Guided Explainable Few-Shot Learning (EGxFSL) with Explainability-Guided Active Learning (xGAL) to tackle both challenges simultaneously.
The EGxFSL framework integrated radiologist-provided regions of interest using a Grad-CAM-based Dice loss, achieving substantial classification improvements (3-20 percentage points) while ensuring clinically aligned attention across MRI and chest X-ray modalities. The complementary xGAL strategy demonstrated strong data efficiency, achieving competitive performance with only 680 labeled samples and delivering 15-38\% relative improvements over random sampling. Grad-CAM visualizations confirmed that guided models consistently attended to diagnostically relevant regions, addressing a key barrier to clinical adoption.
This combination of enhanced interpretability and data efficiency makes the framework particularly suitable for resource-constrained clinical environments, rare disease diagnosis, and specialized imaging protocols. The dual-framework effectively bridges model performance and interpretability, offering a practical pathway for deploying trustworthy AI systems in medical imaging where both accuracy and transparency are essential.

\section*{Acknowledgment}
This work was supported by the National Science Foundation under Grant No. \href{https://www.nsf.gov/awardsearch/showAward?AWD_ID=2346643}{\#2346643}, the U.S. Department of Defense under Award No. \href{https://dtic.dimensions.ai/details/grant/grant.14525543}{\#FA9550-23-1-0495}, the U.S. Department of Education under Grant No. P116Z240151 and National Research Platform (NRP) Nautilus HPC cluster \cite{10.1145/3708035.3736060}.
Any opinions, findings, conclusions or recommendations expressed in this material are those of the author(s) and do not necessarily reflect the views of the National Science Foundation, the U.S. Department of Defense, or the U.S. Department of Education.

\section*{References}
\bibliographystyle{IEEEtran}
\bibliography{bibtext}

\end{document}